\documentclass[11pt]{article}
\usepackage{acl2015}
\usepackage{times}
\usepackage{url}
\usepackage{xcolor}
\usepackage{soul} 
\usepackage{amsmath,amssymb,amsfonts}
\usepackage{graphicx}
\usepackage{xparse,array,booktabs}
\usepackage{makecell}
\usepackage{tikz}
\usepackage{pgfplots}
\usepackage{pgfplotstable}
\usepgfplotslibrary{dateplot}

\pgfplotsset{width=7cm,compat=1.8}
\usepackage{sfmath}

\title{COVID-19: Detecting Depression Signals during Stay-At-Home Period}

\author{ Jean Marie Tshimula,$^{1}$ Belkacem Chikhaoui,$^{1,2}$ Shengrui Wang$^{1}$ \\
  \normalsize $^{1}$Department of Computer Science, Université de Sherbrooke, QC J1K 2R1, Canada \\
  $^{2}$LICEF Research Center, Université TÉLUQ, QC H2S 3L5, Canada\\
 {\tt \{kabj2801,shengrui.wang\}@usherbrooke.ca } \\
 {\tt {belkacem.chikhaoui}@teluq.ca }
 }
 
\date{}

\begin{document}
\maketitle
\begin{abstract}
The new coronavirus outbreak has been officially declared a global pandemic by the World Health Organization. To grapple with the rapid spread of this ongoing pandemic, most countries have banned indoor and outdoor gatherings and ordered their residents to stay home. Given the developing situation with coronavirus, mental health is an important challenge in our society today. In this paper, we discuss the investigation of social media postings to detect signals relevant to depression. To this end, we utilize topic modeling features and a collection of psycholinguistic and mental-well-being attributes to develop statistical models to characterize and facilitate representation of the more subtle aspects of depression. Furthermore, we predict whether signals relevant to depression are likely to grow significantly as time moves forward. Our best classifier yields F-1 scores as high as 0.8 and surpasses the utilized baseline by a considerable margin, 0.173. In closing, we propose several future research avenues.
\end{abstract}

\section{Introduction}
The ongoing coronavirus outbreak has been officially defined a global pandemic by the World Health Organization (WHO) on March 11, 2020. Coronavirus disease 2019 (COVID-19) is an infectious disease caused by a newly discovered coronavirus \cite{WHO:19}. COVID-19 causes a respiratory illness characterized by symptoms such as cough, fever, difficulty breathing, and pneumonia in both lungs. These symptoms may take up to 14 days to appear after exposure to COVID-19. COVID-19 spares no one and infects people of all ages. Older people and those with pre-existing medical conditions like cardiovascular disease, diabetes, chronic respiratory disease, and cancer appear to be more vulnerable to becoming severely ill with COVID-19 \cite{WHO:19,CanadaCovid}.

WHO has reported a drastic increase in confirmed cases and deaths all over the world. To mitigate the rapid spread of COVID-19, many countries have forbidden indoor and outdoor gatherings in excess of particular numbers of people; asked non-essential services, nonprofit entities, and retail businesses to close; issued stay-at-home orders for their residents; and advised them to practice social distancing and avoid all non-essential travel abroad. We are living through a pivotal moment in history. The onslaught of the pandemic has severely challenged our economic systems \cite{McKibbinEconomyCovid19} and caused substantial changes to people's daily routine. The current pandemic can affect people both physically and psychologically \cite{WangCovid19}. For example, in China, 96.2\% of clinically stable COVID-19 patients in the early recovery phase reported significant post-traumatic stress disorder (PTSD) symptoms \cite{Bo2020}. Psychological distress is increasing worldwide and may have long-lasting consequences and repercussions on mental health \cite{Brooks:20,Gunnell2020,Li2020aaa,MengCovid19000a}.

Given the developing situation with the pandemic, social media allows people to inform themselves and get updates from official sources. People may naturally panic when seeing headlines announcing bad news and numbers of cases. This may affect ways in which individuals express themselves and share opinions, thoughts, and personal experiences with others. The emotion and language in social media postings may potentially indicate feelings such as loneliness \cite{Guntuku:2019}, anxiety, anger and stress, among others \cite{DeChoudhury13}. For instance, a person may express emotional reactions that can be unpleasant, disturbing, and overwhelming. Emotional problems like anxiety and depression manifest themselves as feelings of inner emotional distress. Mental health issues can comprise a wide range of disorders that affect mood, thinking, and behavior. Some examples of mental illness include PTSD, depression, anxiety disorders, addictive behaviors, etc. In this paper, our primary interest is in depression. Depression is a serious condition that can cause a persistent feeling of sadness and loss of interest and can affect a person's daily life \cite{Kanter:08}. Survey research conducted by Mental Health Research Canada found that feelings of depression are rising constantly \cite{mhrc2020}. Before the pandemic, 7\% of Canadians reported high levels of depression. This rate has risen to 16\% during the stay-at-home period and 22\% predict high levels of depression if social isolation continues for two more months.

Recognizing early signs of depression is of critical importance and can aid mental health services in assessing the impact of the pandemic on the population and implementing healthier coping strategies to build personal resilience. In addition, appropriate services can be provided for those in need. In this paper, we leverage social media postings to detect signals relevant to depression due to COVID-19. To this end, we build a corpus of postings shared on Twitter during the stay-at-home period. We make use of a topic modeling approach to generate topics addressed by individuals and evaluate language features from topic words to determine whether they indicate signals for depression. It should be noted that we retain solely depression-indicative topics and collect individuals who engage with these topics to investigate their posting histories since the onset of the stay-at-home order. Specifically, this work makes the following contributions:

\begin{itemize}
    \item We demonstrate the effectiveness of our data collection and data pre-processing strategy to gather social media postings containing signals relevant to depression.
    \item We capture evidence from a corpus of postings and potential individuals who manifest signals for depression and consider them as an experimental group. We measure the similarity between different topics addressed by  individuals in the experimental group to discover their overlapping behavioral characteristics and understand their linguistic idiosyncrasies.
    \item We develop models to predict whether signals relevant to depression are likely to grow significantly as time moves forward.
\end{itemize}

\section{Related Work}
The role of social media in mental health has been explored by De Choudhury \cite{DeChoudhury13}. The study suggested a guideline that emphasizes the use of social media postings to gauge what the pertinent mental literature would predict at the individual- and population-levels. This could allow the identification of depressed or otherwise at-risk individuals through the large-scale passive monitoring of social media \cite{Guntuku2017a}. Recently, research has associated social media with several mental health conditions, including stress \cite{Guntuku2019a,Koustuv2017,Thelwal2017}, post-traumatic stress disorder \cite{Glen2014a,Glen2014b,He2012} and depression \cite{Guntuku2017a,Cacheda2019,Glen2015,DeChoudhury22,Jamil2017,Resnik1,Resniklda,Resnik2,Sadeque2018,Schwartz2014,Jia2018,Tsugawa2015}.

To quantify depression from texts, De Choudhury et al. proposed a social media depression index to identify levels of depression among individuals and predict social network behavior changes related to post-partum depression using several features, including structural properties of social networks \cite{DeChoudhury22}. While some studies rely exclusively on open-vocabulary analysis and lexicon-based techniques such as Linguistic Inquiry and Word Count (LIWC) \cite{Pennebaker:15} to build a classifier, other studies couple LIWC with topic modeling features \cite{Resniklda,Stark2012,Tadesse2019,Zhai2012a}. 
For instance, Coppersmith et al. used LIWC to demonstrate characteristic differences in language use for mental disorders \cite{Glen2014a}. Their approach utilizes uni-grams and 5-grams to indicate the presence of mental health conditions. Stark et al.~\shortcite{Stark2012} combined LIWC and latent Dirichlet allocation (LDA)-based features in the classification of social relationships. Resnik et al.~\shortcite{Resniklda} explored the value-add of topic modeling in text analysis for depression and showed that topic models can take us beyond the LIWC categories to relevant themes related to depression and neuroticism as a strongly associated personality measure. Another work of Resnik et al.~\shortcite{Resnik2}  investigated the use of supervised topic models in the analysis of linguistic signals for detecting depression.  Tadesse et al.~\shortcite{Tadesse2019} demonstrated that multiple feature combinations (LIWC+LDA+bi-gram) can yield competitive results. In this paper, we take a step forward by combining LDA with bi-gram, LIWC and other psycholinguistic dictionary-based features to identify depression-indicative topics, in order to facilitate the investigation of signals relevant to depression. The rationale behind the incorporation of additional features is to enrich the model to be able to capture depression-related terms and patterns that may escape the LIWC dictionary. We utilize correlation metrics to compare the performance of the proposed features with other alternative feature combinations. 

\section{Detection of depression signals}

\noindent
\textbf{Dataset during the stay-at-home period.}
All data we obtained is public, posted between 12 March 2020 and 25 May 2020,\footnote{This corresponds to the onset of lockdown and the date on which COVID-19 lockdown restrictions began slowly being relaxed across the country.} and made available from Twitter. Specifically, we extracted tweets bearing the words or hashtags: COVID, coronavirus, \#StayAtHome, or \#StayHome. For privacy and ethical reasons, we avoid displaying personally identifiable information, especially names and pseudonyms. Therefore, we randomly replaced such information to ensure the anonymity and privacy of the data. 

To preprocess the data, we limited our set to Canadian users and removed tweets written in a language other than English or French. Additionally, we discarded redundant tweets, retweets without comments, tweets containing only the keyword (i.e., words or hashtags utilized for extraction), and multimedia such as image and video. We removed links in tweets, but kept emojis, since research has proven that emotions within a text can be expressed through the use of emojis \cite{Hauthal:19}. We used the Python {\it Googletrans}\footnote{https://py-googletrans.readthedocs.io/en/latest/} implementation package to translate tweets from French to English. We removed tweets in which the word COVID or coronavirus occurs simultaneously with the term mental health or depression. We believe that people reacting emotionally may avoid combining the two words in a single tweet when it conveys a personal account. Consequently, we assume that these kinds of tweets are more likely to convey information or warnings about mental health. We eliminated stopwords but kept pronouns.\footnote{I, you, she, he, we and they (see Table \ref{tab3EWERR})} Pronouns reveal information on people's emotional state, thinking, and personality \cite{Pennebaker:15}. Chung and Pennebaker~\shortcite{Chung:07} discovered that individuals susceptible to mental illness such as depression more frequently use first-person pronouns, suggesting higher self-attention focus.

To concentrate exclusively on data containing signals relevant to depression, we quantified different aspects of the language usage and patterns of individuals, using automated methods in order to extract features indicative of depression in tweets.\\

\noindent
\textbf{Dataset before the stay-at-home order.} We replicated and applied the same logic as above to collect tweets posted before the stay-at-home order, that is, from 1 January 2020 to 11 March 2020. In total, we extracted 1,006,941 tweets and 161,327 distinct users, that is, users who had at least five tweets.

\subsection{Feature Design}
\noindent
\textbf{Bi-gram features.} We extracted bi-grams from tweets by leveraging the vectors based on the term frequency-inverse document frequency (TF-IDF) approach \cite{Ramos2003IDF,Tadesse2019}. We used TF-IDF as a statistical measure to evaluate how important a word is to each tweet in the corpus. We convert each tweet into its bag-of-word representation and calculate the TF-IDF value of each word utilizing the standard formula (Equation \ref{eqnTFIDF_m1}).

\begin{equation}
\text{TF-IDF} = (1+\log{n_{w,t}})\times\log{\frac{T}{T_w}}
\label{eqnTFIDF_m1}
\end{equation}

\noindent
where the TF-IDF value of word $w$ in tweet $t$ is the log normalization of the number of times the word occurs in the tweet ($n_{w,t}$) times the inverse log of the number of tweets $T$ and $T_w$ the number of tweets containing word $w$.

\begin{table}[!ht]
\centering
\setlength\tabcolsep{1.2pt}
\caption{Prediction quality for depression, for different feature sets and all combinations, as measured using the Pearson {\it r}. For LIWC features, we consider one feature per category and for LDA features, we take one feature per topic.}\label{tab1}
\begin{tabular}{ l c }
\toprule
\text{Feature set} &  \textit{r} \\
\toprule
\text{LIWC} & {0.286}\\
\text{LIWC+LDA} & {0.342}     \\
\text{LIWC+bi-gram} & {0.313}    \\
\text{LIWC+bi-gram+LDA} & {0.371}  \\
\text{LIWC+PLUS+bi-gram+LDA}& {0.506}  \\
\bottomrule
\end{tabular}
\end{table}

\noindent
\textbf{LIWC features.} The Linguistic Inquiry and Word Count (LIWC) dictionary is a widely used psychometrically validated system for psychology-related analysis of language and word classification \cite{Pennebaker:15}. LIWC includes word categories that have pre-labeled meanings. For each tweet, we calculated the number of observed words, using the LIWC dictionary and focusing on three LIWC categories: linguistic dimensions, psychological processes, and personal concerns. For the psychological processes and personal concerns categories, we utilized all of their subcategories, while for the linguistic dimensions category, we exclusively measured the proportion of first-person pronouns in the tweet. \\

\noindent
\textbf{PLUS features.} We extracted depression-related features from the MRC psycholinguistic database\cite{WIL:19}, the WHO glossary of psychiatric and mental health terms \cite{WHO:94}, and the NRC emotion lexicons \cite{Mohammad:13}. The NRC emotion lexicon is a list of English words and their associations with eight basic emotions ({\it anger, fear, anticipation, trust, surprise, sadness, joy}, and {\it disgust}) and two sentiments ({\it negative} and {\it positive}). MRC provides information about 26 different linguistic properties and includes more than 150,000 words with linguistic and psycholinguistic features of each word. For each tweet, we identified depression-related words using the WHO glossary and verified whether these words fall into the NRC emotion lexicons. Specifically, we discarded all the words that imply ``joy'' as the emotional state. Each MRC feature was computed by averaging the scores of all the depression-related words found in the database. 

\begin{table}
\centering
\setlength\tabcolsep{1.2pt}
\fontsize{9.8pt}{9.8pt}\selectfont

\caption{
Top fifteen words for the first five of the 38 validated depression-indicative topics. 
}\label{tab3EWERR}
\begin{tabular}{cc}
\toprule
\text{Topic} & \text{Words} \\
\toprule

1 & \makecell{limit, alone, bad, I,  bored,\\ hard, when, time, wash,  hand,\\ tired, isolation, abuse, social, paper {}} \\ \toprule

2 & \makecell{feeling, myself, mask, mood, extremely,\\ time, affect, out, crisis, mind,\\ bad, finish, way, I, worse {} } \\ \toprule

3 & \makecell{friends, sleep, I, life, suffer,\\ miss,  shit, always, dull, long,\\ end, back, family, hopeless, change {} } \\ \toprule

4 & \makecell{disgust, hell, freaking, I, enemy,\\ worry,  care, moment, invisible,  difficult,\\ feel, bad, health, home, sick {} } \\ \toprule

5 & \makecell{time, sad, home, close, depressed,\\ hard, move, limited, boring, unhappy,\\ stay, services, weird, feel, park {} } \\
\bottomrule
\end{tabular}

\end{table}

\noindent
\textbf{LDA features.} We utilized LDA \cite{lda2003a} to learn the topics addressed from the tweets. LDA is a probabilistic model that discovers latent topics in a text corpus and can be trained using collapsed Gibbs sampling. A topic is a distribution over a fixed vocabulary. As the parameters of LDA, we set $\alpha$ and $\beta$ to 0.01. All extracted topics were used as features.

\section{Experimental Setup and Results}

\noindent
\textbf{Prediction of depression during the stay-at-home period.} We generated 50 topics overall, of which we especially examined topics containing words related to mental health. To this end, we combined PLUS, bi-gram, and LIWC features to identify topics containing depression-related words. The depression-indicative topics were validated by clinical psychologists. Next, we took users who engaged with the 38 depression-indicative topics (see Table \ref{tab3EWERR}) and collected all tweets of these users from 12 March 2020 to 25 May 2020. We kept users who had at least five tweets and considered these users as an experimental group. In total, we were left with 87,236 distinct users and 857,294 tweets. We performed linear regression with elastic-net regularization to predict depression signals derived from previous features and evaluated the quality of prediction using the Pearson correlation ({\it r}). We stratified the dataset for 10-fold cross-validation to separate our training and testing sets. Table \ref{tab1} shows that all of the feature sets combined (LIWC+PLUS+bi-gram+LDA) produce much stronger correlations ({\it r} = 0.506, {\it p} $<$ 0.001) with depression than other alternative combinations or LIWC alone, and perform reliably well at predicting depression. We report that all correlation coefficients meet ({\it p} $<$ 0.05). We observe that adding PLUS features improves significantly on the results yielded by LIWC+bi-gram+LDA by a considerable margin. It should be noted that Pearson correlations between behavior (such as language use) and psychologically-based features rarely surpass an {\it r} of 0.4 \cite{MeyerPsycho2001}.

\begin{table}[tp]
\centering
\fontsize{9.3pt}{9.3pt}\selectfont
\caption{
Prediction performances over time. Bold font indicates the best result for each feature set.
}\label{tab3}
\begin{tabular}{ l c c c}
\toprule
\text{Feature set} &  \text{SVM} &  \text{LR} &  \text{SVM}\\
\toprule
\text{LIWC} & \textbf{\bf0.629} & {0.611} & {0.623} \\
\text{LIWC+LDA} & {0.652} & {0.647} & {\bf0.654}\\
\text{LIWC+bi-gram+LDA} & {0.706} & {\bf0.718} & {0.715}  \\
\text{LIWC+PLUS+bi-gram+LDA} & {\bf0.802} & {0.800} & {0.780} \\
\bottomrule
\end{tabular}
\end{table}

To make predictions over time for signals relevant to depression, we divided our data (857,294 tweets) into one-week periods. Specifically, we separately derived 50 topics from each subset. We prepared the training set using topics from the first to the penultimate week and took topics from the last week as the test set. We utilized three different classifiers: support vector machine (SVM), logistic regression (LR), and random forest (RF). We trained our classifiers with the three feature sets which achieved the highest Pearson's ({\it r}) results in Table \ref{tab1}: LIWC+LDA, LIWC+bi-gram+LDA, and LIWC+PLUS+bi-gram+LDA. We considered the feature set LIWC itself as a baseline. For SVM, we set the regularization parameter $\lambda$ = 0.0001 and the value $\gamma$ of the radial basis function kernel to 0.5 and for RF, we set the number of trees to 500 and the maximum depth and number of features to 3 and 30, respectively. The prediction performances are reported as F-1 scores, i.e., the harmonic mean of precision and recall. 

Table \ref{tab3} shows the results for depression prediction over time. We see that the F-1 scores achieved with SVM, LR, and RF over the used feature sets are significantly higher than 0.5. We observe that SVM yielded the best performance over LIWC+PLUS+bi-gram+LDA features (0.802), surpassing the baseline (0.629) with a substantial improvement of 0.173. We note that the smallest result achieved with LIWC+PLUS+bi-gram+LDA (0.780) is superior to the performance of our second-best features, LIWC+bi-gram+LDA (0.718). These results indicate that LIWC+PLUS+bi-gram+LDA can detect signals relevant to depression more effectively than other features. LIWC+bi-gram+LDA features resulted in better results than LIWC features alone (0.629) or the combination of LIWC and LDA (0.654). We note that prediction quality depends heavily on complementary features, that is, the more a combination includes several features, the more it yields significantly better results.

\begin{table}[tp]
\setlength\tabcolsep{1.2pt}
\fontsize{9pt}{9pt}\selectfont
\centering 
\caption{Similarity between different depression-related topics addressed by individuals between before and during the stay-at-home period.
}\label{tabXXSimilarity}
\begin{tabular}{lccc}
\toprule
    & Similarity& Before & During \\ 
\toprule
\text{LIWC+LDA} & JS & 0.005 & 0.327 \\ 
                         & KL & 0.017 & 0.403 \\ \toprule

\text{LIWC+bi-gram+LDA} & JS & 0.022 & 0.341 \\
                         & KL & 0.02 & 0.335 \\ \toprule

\text{LIWC+PLUS+bi-gram+LDA} & JS & 0.025 & 0.478 \\ 
                         & KL & 0.027 & 0.290 \\
\bottomrule
\end{tabular}
\end{table}

\begin{equation}
\text{KL}(P{\parallel}Q)=\sum_{i\in[n]} p_i\times\log{(\frac{p_i}{q_i})}
\label{eqnKL_m1}
\end{equation}
\begin{equation}
\text{JS}(P{\parallel}Q)= \frac{1}{2}\text{KL}(P{\parallel}M) + \frac{1}{2}\text{KL}(Q{\parallel}M)
\label{eqnJS_m1}
\end{equation}

\begin{figure*}[ht]
    \centering
    \includegraphics[width=130mm]{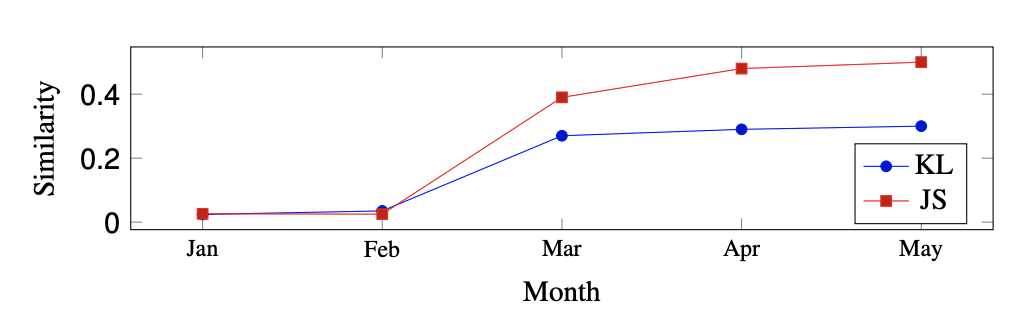}
    \caption{{\color{black}Monthly trends of similarity between depression-related topics addressed by individuals. Note that we utilize LIWC+PLUS+bi-gram+LDA.}}
    \label{FigSIMRiTy813bNSdEh}
\end{figure*}

\begin{figure*}
    \centering
\begin{tikzpicture}
  \centering
  \begin{axis}[
        ybar, axis on top,
        height=8cm, width=14.5cm,
        bar width=0.3cm,
        ymajorgrids, tick align=inside,
        major grid style={draw=white},
        enlarge y limits={value=.1,upper},
        ymin=150, 
        axis x line*=bottom,
        axis y line*=left,
        y axis line style={opacity=0},
        tickwidth=0pt,
        enlarge x limits=true,
        legend style={
            at={(0.5,-0.2)},
            anchor=north,
            legend columns=-1,
            /tikz/every even column/.append style={column sep=0.5cm}
        },
        ylabel={Number of individuals in the topics of interest},
        xlabel={Month (Weeks)},
        symbolic x coords={
           \text{Jan (W1,2,3,4,5)},
           \text{Feb (W1,2,3,4,5)},
           \text{Mar (W1,2,3,4,5)},
           \text{Apr (W1,2,3,4,5)},
           \text{May (W1,2,3,4,5)}
          },
       xtick=data,
      nodes near coords,
      every node near coord/.append style={
            anchor=mid west,
            rotate=90}]
    ]
    \addplot [draw=none, fill=blue!30] coordinates {
      (\text{Jan (W1,2,3,4,5)},4355)
      (\text{Feb (W1,2,3,4,5)},3700)
      (\text{Mar (W1,2,3,4,5)},8861) 
      (\text{Apr (W1,2,3,4,5)},29027)
      (\text{May (W1,2,3,4,5)},9466)};
   \addplot [draw=none,fill=red!30] coordinates {
      (\text{Jan (W1,2,3,4,5)},7600) 
      (\text{Feb (W1,2,3,4,5)},8356)
      (\text{Mar (W1,2,3,4,5)},17679) 
      (\text{Apr (W1,2,3,4,5)},33482)
      (\text{May (W1,2,3,4,5)},37227)};
   \addplot [draw=none, fill=green!30] coordinates {
      (\text{Jan (W1,2,3,4,5)},7946)
      (\text{Feb (W1,2,3,4,5)},8339)
      (\text{Mar (W1,2,3,4,5)},24742) 
      (\text{Apr (W1,2,3,4,5)},35208) 
      (\text{May (W1,2,3,4,5)},23104)};
      
    \addplot [draw=none, fill=orange!30] coordinates {
      (\text{Jan (W1,2,3,4,5)},8111)
      (\text{Feb (W1,2,3,4,5)},8425)
      (\text{Mar (W1,2,3,4,5)},26813) 
      (\text{Apr (W1,2,3,4,5)},35679)
      (\text{May (W1,2,3,4,5)},19451)};
      
      \addplot [draw=none, fill=gray!30] coordinates {
      (\text{Jan (W1,2,3,4,5)},8129)
      (\text{Feb (W1,2,3,4,5)},8509)
      (\text{Mar (W1,2,3,4,5)},9348) 
      (\text{Apr (W1,2,3,4,5)},29195)
      (\text{May (W1,2,3,4,5)},8293)};
  \end{axis}
  \end{tikzpicture}
    \caption{\color{black}The number of individuals who have participated in depression-related topics. We make a weekly count of these individuals in the months before and during the stay-at-home order. For instance, the blue bar in Jan (January) is associated with the first week (W1), the red bar with the second week (W2), and so on.}
    \label{FigXXCOVID19E0AZUIL}
\end{figure*}
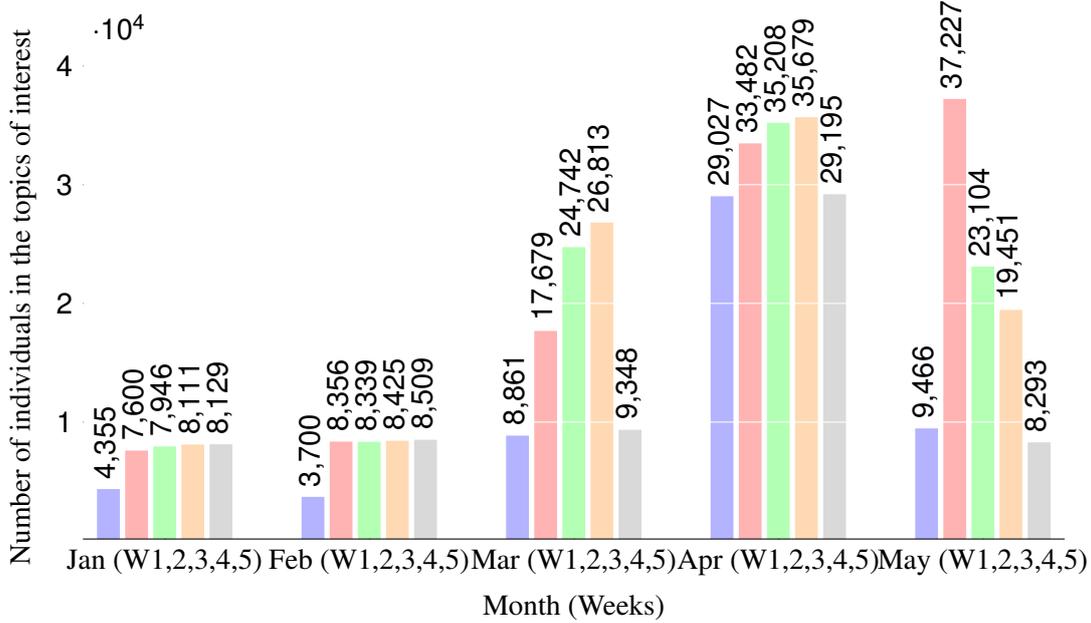

\noindent
\textbf{Similarity between topics before and during stay-at-home restrictions.} To discover overlapping behavioral characteristics of depression-related terms, we experimented with 50 topics on each one-week subset of the data as divided above. Each topic was represented by the top fifteen highest-probability words, out of which we retained solely the top ten depression-related words. We computed topic similarity using measures based on topic word probability distributions \cite{Aletras2014a} (such as Kullback-Leibler divergence (KL) \cite{KL1951}) and topic word sets \cite{Mika2018a} (such as Jaccard similarity (JS) \cite{Jaccard1912}).

Let us look at two discrete probability distributions $P=\{p_i\}_{i\in[n]}$ and $Q=\{q_i\}_{i\in[n]}$ supported on $[n]$. KL measures the difference between two probability distributions (Equation \ref{eqnKL_m1}). Equation \ref{eqnKL_m1} determines how the $Q$ distribution is different from the $P$ distribution. KL is a non-negative, asymmetric distance (i.e., $\text{KL}(P{\parallel}Q)\neq\text{KL}(Q{\parallel}P)$) which yields zero if the two distributions are identical and can potentially equal infinity \cite{Shlens2014bX}. For JS, we measured the similarity between all possible topic pairs. {JS} is a symmetrized, smoothed version of {KL} which measures the total {KL} divergence from the average mixture distribution, $M=\frac{(P+Q)}{2}$ (Equation \ref{eqnJS_m1}). Some salient features of {JS} are that it is always deﬁned, bounded and symmetric, and only vanishes when $P=Q$. When all the top words of a pair of topics are different, JS may result in 0. We found that some topic pairs bear words that include different spellings but are synonyms. To harmonize topic pairs that fall into that situation, we manually replaced synonyms with a single word on either side. We calculated the average JS and KL yielded from different time periods and found that depression-related words were overlapping from one topic to another during the stay-at-home period, and were slightly overlapping before the stay-at-home order (see Table \ref{tabXXSimilarity}).

The Spearman correlation ($\rho$) between the two-similarity metrics is presented. We obtain $\rho$ = 0.839 for LIWC+LDA, $\rho$ = 0.873 for LIWC+bi-gram+LDA, and {$\rho$ = 0.930} for LIWC+PLUS+bi-gram+LDA during the stay-at-home period; and $\rho$ = 0.011 for LIWC+LDA, $\rho$ = 0.016 for LIWC+bi-gram+LDA, and {$\rho$ = 0.02} for LIWC+PLUS+bi-gram+LDA before the stay-at-home order. We report that all correlations are statistically significant ({\it p} $<$ 0.001) and superior to 0.820 during the stay-at-home; and all correlations are not significant before the stay-at-home order ({\it p} $>$ 0.05). {\color{black}In Figure \ref{FigSIMRiTy813bNSdEh}, we utilize LIWC+PLUS+bi-gram+LDA. It should be recalled that the stay-at-home was issued on March 12. Consequently, we combine all the data of March to measure the similarity. Specifically, January and February are fully comprised in the data before the stay-at-home. We obtain a KL of 0.024 and 0.035 in January and February ({\it p} $>$ 0.05), respectively; 0.29 and 0.3 in April and May ({\it p} $<$ 0.001), respectively; and 0.27 in March ({\it p} $<$ 0.05). We get a JS of 0.026 and 0.0249 in January and February ({\it p} $>$ 0.05), respectively; 0.48 and 0.5 in April and May ({\it p} $<$ 0.001), respectively; and 0.39 in March ({\it p} $<$ 0.05)}.

These results indicate strong and meaningful correlations between depression-indicative topics addressed during the stay-at-home. The language in these topics appears to be somewhat similar and recurs from one period to another during the stay-at-home period. This suggests that we should give more attention to this vocabulary when predicting depression from the individual-level. 

{\color{black}Figure \ref{FigXXCOVID19E0AZUIL} shows the trend of individuals who have participated in depression-related topics. We observe a rise of participants within the second week of March, which symbolizes the onset of lockdown; and we note that the number substantially decreased within the fifth week of May, which represents the date on which COVID-19 lockdown restrictions began slowly being relaxed across the country. We calculated the percentage that individuals who have participated in depression-related topics represents to the overall number of individuals collected for each month. We found that 6.9\%, 7.7\%, 28.4\%, 36.4\% and 30.1\%, respectively, for January, February, March, April and May.}

\section{Conclusion}
This study focuses on detecting depression from social media postings, computes the language similarity  between all possible topic pairs addressed by individuals, and predicts the evolution of depression over time. Our best classifier achieves F-1 scores as high as 0.8, which is a 0.173 relative the improvement over the baseline features. The proposed features yield a higher Pearson correlation ({\it r} = 0.506) than other alternative feature combinations and the improvement is statistically significant ({\it p} $<$ 0.001). Prior work found that Pearson correlations between language use and psychologically-based features rarely exceed a value of {\it r} = 0.4, while our result has surpassed this value by 0.106. We measure the similarity between different topics addressed by individuals to discover overlapping behavioral characteristics of depression-related words. We report that the Spearman correlations for this task are statistically significant for all the features utilized, and the proposed features specifically achieve the strongest Spearman correlation. In future work, we aim to include socioeconomic and demographic attributes with network and language information to predict depression at the regional level. Additionally, we would like to investigate affinity relationships between individuals who manifest signs of depression \cite{bX19ieoaiXERTWY,bX19ieoaiXERTWYV222}.

\end{document}